\RequirePackage{amsmath}
\RequirePackage{flexisym}
\RequirePackage[superscript,biblabel,nomove]{cite}
\RequirePackage{graphicx}
\documentclass[12pt,aasms4]{article}
\usepackage{amsmath}
\usepackage{amssymb}
\usepackage{setspace}
\def\gsim{\;\rlap{\lower 2.5pt \hbox{$\sim$}}\raise 1.5pt\hbox{$>$}\;}
\def\lsim{\;\rlap{\lower 2.5pt  \hbox{$\sim$}}\raise 1.5pt\hbox{$<$}\;}
\newcommand\beq{\begin{equation}}
\newcommand\eeq{\end{equation}}

\begin{document}

\Large
\centerline{\bf Reply to Desch et al. (2021) on:}
\centerline{\bf \textit{“Breakup of a long-period comet as}}
\centerline{\bf \textit{the origin of the dinosaur extinction"}}

\medskip
\medskip
\normalsize
\centerline{ Amir Siraj$^{1\star}$ \& Abraham Loeb$^1$}
\medskip
\medskip
\medskip
\centerline{\it $^1$Department of Astronomy, Harvard University}
\centerline{\it 60 Garden Street, Cambridge, MA 02138, USA}
\medskip
\medskip
\small
\centerline{$^{\star}$Correspondence to amir.siraj@cfa.harvard.edu}
\normalsize
\vskip 0.2in
\hrule
\vskip 0.2in

\begin{doublespace}
\section*{Abstract}
{\bf We reply to criticisms by Desch et al. (2021)\cite{2021arXiv210508768D} regarding our \textit{Scientific Reports} paper\cite{2021NatSR..11.3803S}, “Breakup of a long-period comet as the origin of the dinosaur extinction." The background impact rates of main-belt asteroids and long-period comets have been previously dismissed as being too low to explain the Chicxulub impact event. Our work demonstrates that a fraction of long-period comets are tidally disrupted after passing close to the Sun, each producing a collection of smaller fragments that cross the orbit of Earth. This population greatly increases the impact rate of long-period comets capable of producing Chicxulub, making it consistent with the age of the Chicxulub impact crater. Our results are subject to an uncertainty in the number of fragments produced in the breakup event that leaves the conclusion of a cometary origin being more likely than an asteroidal one unchanged.}

\newpage

In a recent commentary, Desch et al. (2021)\cite{2021arXiv210508768D} claim discrepancies in the impact rate of Chicxulub-sized fragments from the breakup of long-period comets near the Sun owing to the abundance of carbonaceous chondrite impactors and the number of fragments produced in tidal disruption events, arguing that the combined reductions in the relative cometary impact rate exceed an order of magnitude, making the Chicxulub impactor less likely to be a comet than an asteroid.

Desch et al. (2021)\cite{2021arXiv210508768D} argue that $30\% - 50\%$ of possible asteroid impactors have carbonaceous chondrite compositions, representing a factor of $3 - 5$ reduction in the relative cometary impact rate compared to the value of 10\% adopted in our model\cite{2021NatSR..11.3803S}. This claim is inadequate for the following reasons.

The 50\% claim is attributed to Morbidelli et al. (2020)\cite{2020Icar..34013631M}. However, Morbidelli et al. (2020) only argues that 50\% of near-Earth objects (NEOs) have low albedos -- not that they have carbonaceous chondrite compositions.

The 30\% claim is attributed to Bottke et al. (2007)\cite{2007Natur.449...48B}. However, the argument made in the paper is actually that 30\% of possible Chicxulub impactors are C-type, X-type, or comet taxonomy NEOs \cite{2007Natur.449...48B}. The clearest way of realizing that not all of these possible asteroid impactors could have carbonaceous chondrite compositions is by looking at the observational evidence. Namely, one must consider the fact that only 5\% of observed meteorite falls in Antarctica (only 4\% globally) have carbonaceous chondrite compositions \cite{2006mess.book..869Z}, which is an order of magnitude lower than Desch et al. (2021)\cite{2021arXiv210508768D} claim the fraction of possible asteroid impactors with carbonaceous chondrite compositions to be. There is no clear reason for some chondrites to be systematically favored over others in meteorite fall surveys. Nonetheless, our model conservatively assumes that the fraction of carbonaceous chondrites is a factor of two higher than observed in meteorite falls, and additionally, in the following paragraph, we consider a calibration of the carbonaceous chondrite fraction that is unrelated to the proportion of ordinary chondrites.

A separate way of reaching the same conclusion, in a manner independent of the proportion of ordinary chondrites observed in meteorite falls, is by considering the fact that X-type asteroids include E-type (Enstatite) and M-type (metallic) asteroids. The relative abundance of carbonaceous meteorite falls relative to carbonaceous, enstatite, and metallic meteorite falls are 35\% on Antarctica (and only 20\% globally), which are both even more punitive to the asteroid scenario than the conservative proportion we adopted in our model (40\%). Therefore, a factor of $3 - 5$ (corresponding to a claim of $30\% - 50\%$ of possible asteroid impactors having carbonaceous chondrite compositions) is inconsistent with observational data, which point to a fraction in the range $5\%-10\%$ (from which we conservatively adopted the upper bound). A Chicxulub-size asteroid with a carbonaceous chondrite composition should only strike Earth once per $\sim 3.5 \; \mathrm{Gyr}$, which represents a rate too low to explain the observed impact. On the other hand, our model\cite{2021NatSR..11.3803S} explains the timing of the impact in a manner consistent with observational evidence.

The only cometary sample-return mission to date, \textit{Stardust}, found that Comet 81P/Wild 2 had a carbonaceous chondritic composition, with the chondrules bearing resemblance to those from asteroids\cite{2008Sci...321.1664N, 2012E&PSL.341..186B}. Specifically, the sample exhibited similarities to the CM and CI sub-groups \cite{2008M&PS...43..261Z, 2011PNAS..10819171C}, contrary to the assertion from Desch et al. (2021)\cite{2021arXiv210508768D} that only CI compositions are consistent with comets. This reinforces the conservative nature of our comparison of the abundance of carbonaceous chondrites overall, rather than specific sub-groups, since focusing on CM would be even more restrictive for the asteroid scenario. Carbonaceous chondrite compositions could be an order of magnitude more abundant in comets than in asteroids; future observations of comets will calibrate this proportion. 

Desch et al. (2021) argue that the number of fragments produced per tidal disruption event, $N \sim 20$, rather than $N \sim 630$, which was adopted in our paper. This is possible only as a limiting case, since it assumes purely gravitational contraction — not accounting for tidal disruption, outgassing, torque, etc., which are known to act on comets that are being tidally disrupted by the Sun and which lengthen the contraction timescale that sets the number of clumps. The value we adopted is the upper limit ($N \sim t / \tau$)\cite{1998P&SS...46.1677H}. Therefore a range of values should be considered, ranging from purely gravitational contraction to the spreading timescale greatly exceeding the contraction timescale.

However, let us adopt the limiting case, the number of fragments argued by Desch et al. (2021)\cite{2021arXiv210508768D}, $N \sim 20$, and consider the effect on the relative cometary impact rate. The impact rate \cite{2021NatSR..11.3803S} scales as, $1 + (0.2 \times N^{2/3})$. Adopting $N \sim 20$ yields a reduction in impact rate of only a factor of $\sim 7$ relative to the $N \sim 630$ upper limit used in our paper.

In other words, $N \sim 20$ would correspond to Chicxulub-sized fragments for a progenitor comet with diameter $\sim 19 \mathrm{\; km}$. The overall reduction in the impact rate of Chicxulub-sized fragments relative to $N \sim 630$ would only be a factor of $\sim 7$, which does not change the conclusion of our model that comet fragments are dynamically more likely than asteroids to explain Chicxulub. Even adopting the full reduction factor of $\sim 7$ associated with the value of $N$ from Desch et al. (2021)\cite{2021arXiv210508768D}, a cometary Chicxulub impactor would be up to $\sim 2$ times more likely than an asteroid Chicxulub impactor\cite{2021NatSR..11.3803S}.

Finally, let us comment on iridium. The most recent analysis of the global iridium fluence focuses only on sites with reliable and representative data \cite{2013LPI....44.2405M}, and finds that two independent datasets converge on a favored impactor mass of $\sim 10^{17} \; \mathrm{\; g}$, which is the same as the impactor in our model. Indeed, the detailed analysis of the global iridium fluence favors a cometary origin for the Chicxulub impactor.\cite{2013LPI....44.2431M}

In summary, our model still holds, and the implied rate of cometary Chicxulub impactors still likely exceeds that of asteroidal Chicxulub impactors, with an uncertainty of a factor of $\sim 7$ in the impact rate associated with the number of fragments, a factor that does not affect the qualitative conclusion. Our model can be tested by future investigation of the chemical composition of Chicxulub ejecta (Meier, M. et al., in preparation).

\end{doublespace}

\vskip 0.45in
\hrule
\vskip 0.15in

\small
\noindent





\end{document}